\documentstyle[11pt,psfig,newpasp,twoside]{article}
\index{summary}
\index{instructions}
\index{template}
\index{supernova remnants!G320.4--1.2}
\index{supernova remnants!MSH 15--5{\em 6}}
\index{supernova remnants!G296.5+10.0}
\index{neutron stars!1E~1207.4-5209}
\index{pulsars!PSR~B1509-58}
\index{supernova remnants!morphology}
\index{ISM!HI}

\def\HI{H\,{\sc i}}
\def\kms{km~s$^{-1}$~}

\def\edcomment#1{\iffalse\marginpar{\raggedright\sl#1\/}\else\relax\fi}
\reversemarginpar

\begin{document}
\title{Environs of Bilateral Supernova Remnants with Neutron Stars}
 \author{G. Dubner\altaffilmark{1}}
\affil{IAFE, CC 67, Suc. 28, 1428 Buenos Aires, Argentina}
\author{E. Giacani\altaffilmark{1}}
\affil{IAFE, CC 67, Suc. 28, 1428 Buenos Aires, Argentina}
\author{B. M. Gaensler}
\affil{Harvard-Smithsonian Center for Astrophysics, 60 Garden Street,
Cambridge, MA 02138, USA}
\author{W. M. Goss}
\affil{VLA - NRAO, P.O. Box 0, Socorro, NM 87801, USA}
\author{A. Green}
\affil{Astrophysics Dep., School of Physics, Univ. of Sydney, NSW 2006,
Australia}

\altaffiltext{1}{Member of the Carrera del Investigador Cient\'\i fico
of CONICET (Argentina)}
\begin{abstract}

We report on Australia Telescope Compact Array (ATCA) \HI\ observations 
carried out in the direction of bilateral supernova remnants (SNRs) 
with associated neutron stars:
G296.5+10.0 and G320.4--1.2, in a search for the origin of such morphology. 
From these studies we conclude that in the case
of G296.5+10.0, located far from the Galactic plane, the \HI\ distribution 
has not influenced the present morphology of the SNR. In the case of
G320.4--1.2, evolving in a denser medium, the combined action of the
central pulsar, PSR B1509-58, with the peculiar distribution of the
surrounding medium, has determined the observed characteristics of the SNR.
\end{abstract}

\section{Introduction}

Several Galactic supernova remnants (SNRs) exhibit an unusual bilateral
morphology, characterized by a clear axis of symmetry, two bright limbs
on either side and low level of emission near the top and bottom along 
the symmetry axis. The origin of this ``barrel-shaped'' appearance has 
provoked considerable debate for the past few years.
A detailed study of the gaseous environs of bilateral SNRs is a very
useful tool to disentangle intrinsic origins (like the presence of
biconical beams from a central neutron star (NS), asymmetric explosions,
etc.) from extrinsic causes (stratification of the interstellar
density, the strength and orientation of the ambient magnetic field, etc.). 

As a part of an ongoing project to observe the environs of bilateral
SNRs, we have conducted a detailed \HI\ study around G296.5+10.0 and
G320.4--1.2. These two bilobular SNRs share the 
characteristic of harbouring an eccentric X-ray pulsar in the interior
(the source 1E 1207.4--5209 inside G296.5+10.0, Helfand \& Becker 1984, and
PSR B1509-58 associated with G320.4--1.2, Seward \& Harnden 1982). 
Using the 
Australia Telescope Compact Array (ATCA) we surveyed wide fields around these
extended SNRs, looking for peculiar alignments, and/or properties of 
the surrounding gas which may give clues as to the origin  
of the observed radio morphology.

\section{Observations}

Mosaic interferometric $\lambda$ 21 cm \HI\ observations were 
carried out using the Australia
Telescope Compact Array (ATCA), during four sessions of 13 hours each
in 1998.  Short spacing data were added
to the interferometric database in order to recover structures at all
spatial frequencies. The single dish data were taken with the
Argentinian 30-m radiotelescope (IAR) for G296.5+10.0, and with the
Parkes 64-m telescope (from the SGPS survey, McClure-Griffiths et al.
2001) for G320.4--1.2. Table 1 summarizes the observational parameters.

\begin{table}
\caption{Observational parameters}
\begin{tabular}{cccccc} \hline
Source&Observed &Mosaic &Velocity&Beam&Noise\\ 
&field&pointings&resol. (km/s)&(arcmin)&(mJy/beam)\\ \hline
G296.5+10.0&3$^\circ.5 \times 3^\circ.5$&109&0.82
&4.0$ \times 2.7$&53\\
G320.4--01.2&1$^\circ.1 \times 1^\circ.1$&19&0.82 
&4$.0 \times 2.7$&30\\ \hline\hline
\end{tabular}
\end{table}

\section{Results}
\subsection{The SNR G296.5+10.0}
This SNR has a bilateral appearance both in radio and X-rays. Its
symmetry axis is oriented almost perpendicular to the Galactic plane.
The radio quiet neutron star 1E 1207.4-5209
is located about 6$^\prime$ from the geometric center. 

After inspection of the \HI\ emission within the observed velocity
range (--225,+350) \kms, we find that the best morphological
correspondence between the surrounding \HI\ gas and the radio, X-rays
and optical emission associated with G296.5+10.0, occur between V$_{\rm
LSR} \sim -17$ \kms and --14 \kms. In this velocity range some good
matches are observed to the NE and along a short portion of the western
limb of the SNR (Figure~1, left panel).

\begin{figure}
\centerline{\psfig{file=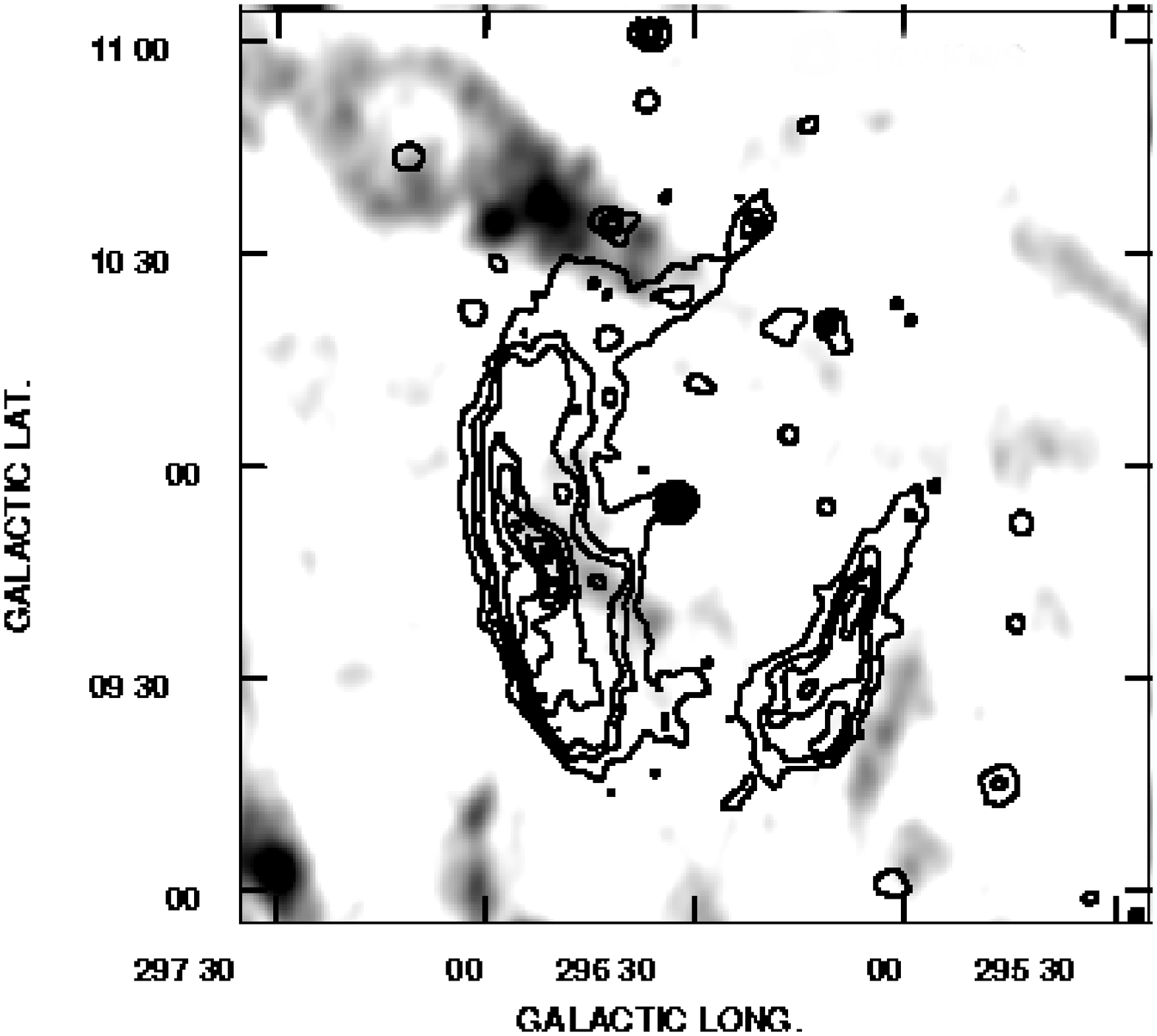,height=6cm}
\hspace{8mm}
\psfig{file=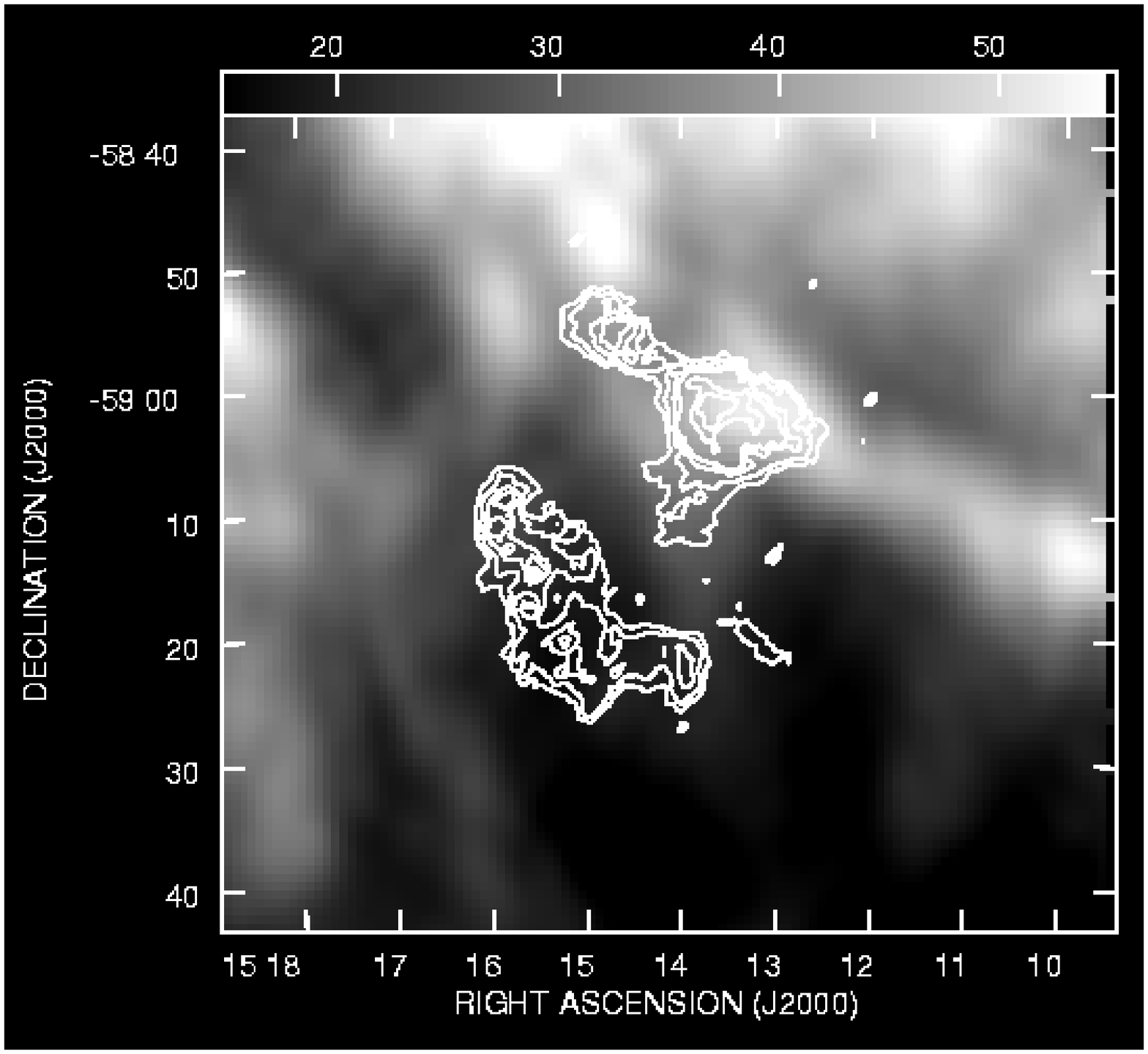,height=6cm}}
\caption{Left panel: Comparison of a ROSAT
X-ray image of G296.5+10.0 (in contours) and the \HI\ distribution
around V$_{\rm LSR} \sim -15$ \kms (in greyscale, from Giacani et al.
2000).  Right panel: Comparison of radio continuum emission of G320.4--1.2 
(white contours, from Gaensler et al. 1999) with the \HI\ distribution 
in the --76 \kms$\leq$ V$_{\rm LSR} \leq$ --66 \kms interval (in
greyscale).}
\label{fig:1}
\end{figure}

The distribution of the absorbing column density integrated between the
observer and the G296.5+10.0 shows a striking hole in the \HI\
emission, exactly coincident with the X-ray source 1E 1207.4-5209 in
all the three coordinates ({\it l}, {\it b} and V$_{\rm LSR}$). From
this fact we independently confirm the association between the NS and
the SNR. The X-ray flux from the radio quiet NS must be heating
the local gas around 1E 1207.4-5209, thus providing a hot background 
against which the emission of the cold foreground \HI\ is self-absorbed.

\subsection{The SNR G320.4--1.2}

This SNR has a bilateral appearance in the radio range. The 
optical/radio/X-ray nebula RCW 89 is located on the NW extreme of
G320.4--1.2. The young pulsar PSR B1509--58 lies between the two main
radio components. 
In the X-ray wavelengths  the system is dominated by the pulsar
emission (also detected at radio wavelengths and in $\gamma-$rays).
Recent {\it Chandra} images show with great detail the bright 
X-ray synchrotron nebula associated with the pulsar, with a collimated 
feature that is interpreted as a relativistic jet directed along the 
pulsar spin axis (Gaensler et al. 2002, this proceedings).

The \HI\ survey of the environs have shown that G320.4--1.2 evolves
within an elongated cavity, with the NW radio lobe interacting with a
dense wall (atomic density
$n \sim 12-15$ cm$^{-3}$)  (Figure~1, right panel). Such 
wall would be responsible for the flat appearance of the NW half of the SNR.
The interaction between the collimated relativistic outflow from the
pulsar and the densest clump of this \HI\ feature near RCW 89, would be
responsible for the formation of the bright radio and X-rays knots and
for the H$_\alpha$ emission of the nebula. To the SE, the SNR expands 
into a lower
density medium, thus explaining the farther distance attained by this
lobe, as well as its semi-circular morphology.  
An extended version of this work is presented in Dubner et al. (2002).

\section {Conclusions}
 
Based on a detailed study of the distribution and kinematics of the \HI\ 
around the SNRs G296.5+10.0 and G320.4--1.2, we conclude: {\it (a)} in the 
case of G296.5+10.0, which is placed far from the Galactic plane, the 
surrounding gas has had little influence in the morphology of the SNR. Our
observations have confirmed the physical association between this SNR
and the isolated radio-quiet neutron star 1E 1207.4-5209. Asymmetric
explosion, and/or biconical outflows from the neutron star, must have
contributed to the present bilateral morphology of G296.5+10.0.;
{\it (b)} in the case of G320.4--1.2, located in a higher density
environment, the properties of the surrounding medium combined with the
characteristics of the associated pulsar have conditioned the present
morphology of the  SNR.
The \HI\ observations have revealed the existence of a cavity in the
interstellar medium. We propose that as a consequence of the interaction 
of the expanding SN shock with the northern wall of this cavity,  the NW 
radio lobe of G320.4--1.2 acquired a flatten appearance. Also, the encounter 
of the relativistic pulsar outflow with this wall is
probably responsible for the formation of the optical nebula RCW 89.

\acknowledgments

G.D. and E.G. are grateful to the organizers of the Workshop for the financial
support to facilitate their attendance to the same. This research was
partially funded through a Cooperative Research Program between CONICET
(Argentina) and the NSF (USA). The ATCA is funded by the Commonwealth
of Australia for operation as National facility, managed by CSIRO.
The NRAO is a facility of the NSF operated under cooperative agreement by AUI.

\end{document}